\documentclass[showpacs,floats,twocolumn,prb,floatfix]{revtex4}
\usepackage{amsmath}
\usepackage{amssymb}
\usepackage{bm}
\usepackage{graphicx,epsfig,color}

\begin{document}

\title{Localization properties of one-dimensional
Frenkel excitons:\\ Gaussian versus Lorentzian diagonal disorder}

\author{S. M. Vlaming}
\affiliation{Centre for Theoretical Physics and Zernike Institute
for Advanced Materials, University of Groningen, Nijenborgh 4, 9747
AG Groningen, The Netherlands}

\author{V. A. Malyshev}
\affiliation{Centre for Theoretical Physics and Zernike Institute
for Advanced Materials, University of Groningen, Nijenborgh 4, 9747
AG Groningen, The Netherlands}

\author{J.\ Knoester}
\affiliation{Centre for Theoretical Physics and Zernike Institute
for Advanced Materials, University of Groningen, Nijenborgh 4, 9747
AG Groningen, The Netherlands}

\date{\today}

\begin{abstract}

We compare localization properties of one-dimensional Frenkel
excitons with Gaussian and Lorentzian uncorrelated diagonal
disorder. We focus on the states of the Lifshits tail, which
dominate the optical response and low-temperature energy transport
in molecular J-aggregates. The absence of exchange narrowing in
chains with Lorentzian disorder is shown to manifest itself in the
disorder scaling of the localization length distribution. Also, we
show that the local exciton level structure of the Lifshits tail
differs substantially for these two types of disorder: In addition
to the singlets and doublets of localized states near the bare band
edge, strongly resembling those found for Gaussian disorder, for
Lorentzian disorder two other types of states are found in this
energy region as well, namely multiplets of three or four states
localized on the same chain segment and isolated states localized on
short segments. Finally, below the Lifshits tail, Lorentzian
disorder induces strongly localized exciton states, centered around
low energy sites, with localization properties that strongly depend
on energy. For Gaussian disorder with a magnitude that does not
exceed the exciton bandwidth, the likelihood to find such very deep
states is exponentially small.
\end{abstract}


\maketitle

\section{Introduction}
    \label{Introduction}

The term "exciton", introduced seventy-five years ago in the
pioneering works of Frenkel~\cite{Frenkel32} and
Wannier,~\cite{Wannier37} has become widely used to explain optical
and transport properties of a large variety of organic and
semiconductor materials.~\cite{Knox63,Davydov71,Agranovich82} Within
this general context, low-dimensional (nanoscale) systems currently
attract particular attention.\cite{Scholes06}

In low-dimensional systems, an important factor influencing the
exciton states is the presence of disorder, which results from
static fluctuations in the host, different growth conditions, as
well as imperfections of the systems themselves. Disorder induces
localization of the exciton states~\cite{Anderson58,Kramer93} on
certain segments of the system; the linear extent of such segments is
usually referred to as the localization length. These localized
states consequently lead to the appearance of tails in the density
of states (DOS) outside of the bare exciton band, which are commonly
known as Lifshits tails.~\cite{Halperin66} These states play a key
role in the low-temperature energy transport as well as in the
optical properties of a wide spectrum of materials, such as
conjugated oligomer aggregates~\cite{Spano06} and
polymers,~\cite{Hadzii99} molecular J-aggregates,~\cite{Kobayashi96}
semiconductor quantum wells and quantum dots,~\cite{Takagahara03}
semiconductor quantum wires,~\cite{Akiyama98} as well as
photosynthetic light harvesting complexes~\cite{vanAmerongen00} and
proteins.~\cite{Berlin07}

In discrete materials, characterized by interacting sites (molecular
aggregates, conjugated polymers, photosynthetic complexes, spin
systems), various types of disorder may be considered. Commonly used
are uncorrelated diagonal (on-site) and off-diagonal disorder, where
different choices still can be made for the types of disorder
distributions; a Gaussian distribution and a box-like distribution
are the most common choices. An interesting alternative is provided
by a Lorentzian distribution, because the tight-binding model with
diagonal uncorrelated Lorentzian disorder (known as the Lloyd model)
is one of the few disorder models that allows for analytical
calculation of several physical quantities, such as the averaged
one-particle Green's function~\cite{Lloyd69} and the variance of the
Lyapunov exponent, which is a measure of the localization
length.~\cite{Deych00} The details of the disorder model affect the
optical response and transport properties of the above mentioned
systems, sometimes substantially. While in comparison to Gaussian
and box-like distributions the choice of Lorentzian disorder is not
very common, it is worth noting that the latter naturally occurs in
random systems dominated by dipolar interactions (see Appendix
\ref{non Gaussian disorder}).

In this paper, we perform a comparative study of uncorrelated
Gaussian and Lorentzian diagonal disorder in one-dimensional
excitonic systems, with particular interest in the localization
properties, level structure, and statistics of the wave functions of
the exciton states in the Lifshits tail. We will only consider
moderate disorder magnitudes, where the exciton states still
correspond to electronic excited states that are coherently shared
by a number of molecules. There is an important difference between
Gaussian and Lorentzian disorder: the former is characterized by a
bounded second moment, while the second moment of the latter
diverges. Distributions with a finite second moment give rise to
exchange narrowing:~\cite{Knapp84} because the exciton wave
functions are coherently shared by a number ($N^*$) of monomers,
they feel an effective disorder of magnitude $\sqrt{N^*}$ times
smaller than the bare value. This effect explains the narrowness of
the optical spectra of molecular J-aggregates as compared to their
monomeric counterparts.~\cite{Knapp84} Exchange narrowing does not
occur for Lorentzian disorder.~\cite{Eisfeld06} As we will show,
this difference strongly affects the disorder scaling of the
localization properties of the excitons, resulting in differences in
the optical and transport properties. In addition, for Lorentzian
disorder the relatively high density of sites with a very low energy
(well outside of the exciton band), also plays an important role in
the exciton optical dynamics.

The outline of this paper is as follows. In the next section, we
present our model and analyze the exchange narrowing effect.
Section~\ref{sec:wavefunctions} contains the results of numerical
simulations and discusses the local level structure and the
statistics of the wave functions in the Lifshits tail for Lorentzian
disorder, which we compare to previously obtained results for
Gaussian disorder. In Sec.~\ref{sec:scaling}, we discuss the scaling
properties of the average as well as the standard deviation of the
localization length in the Lifshits tail for both types of disorder.
Section~\ref{sec:summary} summarizes the paper. In Appendix~\ref{non
Gaussian disorder} we show that systems with random dipolar
interactions provide physical realizations of the Lorentzian
disorder model that is analyzed in this paper. Finally, in Appendix
\ref{Exchange narrowing}, we present some mathematical details of
derivations outlined in Sec. \ref{subsec:exchange narrowing}.

\section{Theoretical background}
\label{sec:background}

\subsection{Frenkel Hamiltonian}
\label{sec:hamiltonian}

We consider a disordered Frenkel exciton chain of $N$ molecules,
described by the Hamiltonian
\begin{equation}
\label{ham}
    \hat{H} = \sum_{n=1}^N E_n \left|n\rangle \langle n\right| -
J \sum_{n=1}^{N-1} \left(\left|n\rangle \langle n+1
\right|+\left|n+1\rangle \langle n \right|\right) \ .
\end{equation}
Here $|n \rangle$ denotes the state in which molecule $n$ is
excited, while all other molecules are in their ground state. $E_n$
denote the molecular excitation energies and $-J~(J>0)$ is the
nearest-neighbor interaction. We neglect interactions beyond the
nearest-neighbor, as this allows for an analytical discussion of
several important quantities. In Sec.~\ref{sec:scaling} we will
briefly comment on extension of the model to include long-range
dipole-dipole interactions. We account for disorder by including a
stochastic component in the site energies $E_n$. Two distributions
of $E_n$ will be considered, a Gaussian and a Lorentzian, both with
zero mean (the transition energy of an isolated molecule is set to
zero)
\begin{equation}
\label{Gauss}
    G(E_n)=\frac{1}{\sqrt{2\pi}\Delta_G}\
\exp\left(-\frac{E_n^2}{2\Delta_G^2}\right) \ ,
\end{equation}
\begin{subequations}
\begin{equation}
\label{Lorentz}
    L(E_n)=\frac{1}{\pi}\frac{\Delta_L}{E_n^2+\Delta_L^2},
\end{equation}
\end{subequations}
where $\Delta_G$ and $\Delta_L$ denote the standard deviation and
half width at half maximum (HWHM), respectively, which are measures
of the disorder strength.

For a given disorder realization, diagonalizing the Hamiltonian
(\ref{ham}) yields the $N$ exciton wave functions $|\nu \rangle =
\sum_{n=1}^N \varphi_{\nu n} |n \rangle$ and the corresponding
energies $E_{\nu}$. For nonzero disorder strength, these exciton
states will be localized on a length scale that depends on the
energy. The linear optical response is dominated by the states with
a large transition dipole to the ground state, i.e. with a large
oscillator strength $O_{\nu}= \left(\sum_{n=1}^N \varphi_{\nu
n}\right)^2$, where we have assumed that all molecular transition
dipoles are equal, and the oscillator strength of a monomer is set
to unity. These states occur in the neighborhood of the lower
exciton band edge for the disorder-free system, $E_b=-2J$, mostly in
the Lifshits tail, i.e., just below $E_b$. Their typical extension
(localization length) is indicated by $N^{*}$. This quantity can
also be interpreted as the typical number of coherently bound
molecules participating in a particular exciton state.

There are various measures for the localization length of a
particular exciton state. We will consider the one based on the
inverse participation ratio~\cite{Thouless74,Schreiber82,Fidder91}
or participation number. The latter is defined as
\begin{equation}
\label{part}
    N_{\nu} = \left( \sum_{n=1}^N \varphi_{\nu n}^4 \right)^{-1}\ ,
\end{equation}
which reflects the number of molecules that contribute to the
exciton state $\left|\nu \right>$.

\subsection{Exchange narrowing}
\label{subsec:exchange narrowing}

A common property of systems of interacting molecules is that their
delocalized excited states give rise to much narrower spectral peaks
than an ensemble of noninteracting molecules. This phenomenon is
referred to as exchange narrowing.~\cite{Knapp84} Its origin lies in
the fact that the delocalized excited states do not feel the local
disorder magnitude, but rather an average over its variations, which
leads to a reduced effective disorder. More specifically, for an
exciton state spread over a localization segment of length $N^{*}$,
the residual disorder strength is $\sigma=\Delta/\sqrt{N^{*}}$,
where $\Delta$ is the standard deviation ($\Delta^2$ is the second
moment of the disorder distribution).

For Lorentzian disorder, exchange narrowing does not occur because
the second moment diverges, which results in the absence of the
exchange narrowing effect.~\cite{Eisfeld06} Below, we briefly sketch
these arguments. For this purpose, we will use the
Hamiltonian~(\ref{ham}) on the basis of the exciton wave functions
of a disorder-free linear chain,
\begin{equation}
\label{n via nu}
    |\nu \rangle = \left( \frac{2}{N+1} \right)^{1/2} \sum_{n=1}^N
    \sin \left( \frac{\pi \nu n}{N+1} \right) |n \rangle \ ,
\end{equation}
with $\nu=1,2,...,N$. This yields
\begin{subequations}
\begin{equation}
\label{ham ex}
    \hat{H} = \sum_{\nu=1}^N E_{\nu} |\nu \rangle \langle \nu| +
    \sum_{\nu\nu^{\prime}=1}^N H_{\nu\nu^{\prime}} |\nu \rangle
    \langle \nu^{\prime}| \ ,
\end{equation}
with
\begin{equation}
\label{Enu}
    E_{\nu} = -2J \cos\left( \frac{\pi \nu}{N+1} \right) \ ,
\end{equation}
and
\begin{equation}
\label{Hnunu'}
    H_{\nu\nu^{\prime}} = \frac{2}{N+1}
    \sum_{n=1}^N E_n \sin\left( \frac{\pi \nu n}{N+1} \right)
    \sin\left( \frac{\pi \nu^{\prime} n}{N+1} \right) \ .
\end{equation}
\end{subequations}
$H_{\nu\nu^{\prime}}$ is a stochastic matrix fluctuating from one
realization of the disorder to another. Its diagonal elements
$H_{\nu\nu}$ describe fluctuations of the exciton energies due to
disorder, while the off-diagonal part describes the scattering of
excitons between different states, which eventually results in their
localization. We are interested in the distribution functions $P(H)$
of these fluctuation matrix elements.

For Gaussian diagonal disorder, it is given by (see
Appendix~\ref{Exchange narrowing})
\begin{equation}
 \label{P(H) Gaissian}
    P_{\nu\nu^{\prime}}(H) = \frac{1}{\sqrt{2\pi} k_{\nu\nu^{\prime}}\Delta_G}
    \exp \left(- \frac{H^2}{2k_{\nu\nu^{\prime}}^2 \Delta_G^2}\right) \ ,
\end{equation}
where $k_{\nu\nu} = 1/\sqrt{(3/2)(N+1)}$ and $k_{\nu\nu^{\prime}} =
1/\sqrt{N+1}~~(\nu\neq\nu^{\prime})$. Thus, the distribution
function $P_{\nu\nu^{\prime}}$ is also a Gaussian with standard
deviation $k_{\nu\nu} \Delta_G = \Delta_G/\sqrt{(3/2)(N+1)}$ and
$k_{\nu\nu^{\prime}}\Delta_G = \Delta_G/\sqrt{N+1}$ for the diagonal
and off-diagonal elements, respectively. The observed suppression of
the bare disorder magnitude $\Delta_G$ by a factor of $\sqrt{N+1}$
reflects the exchange narrowing effect.~\cite{Knapp84}

By contrast, for Lorentzian diagonal disorder $P(H)$ reads (see, again,
Appendix~\ref{Exchange narrowing})
\begin{subequations}
\begin{equation}
 \label{P(H)calculated}
    P_{\nu\nu^{\prime}}(H) = \frac{1}{\pi} \>
    \frac{k_{\nu\nu^{\prime}} \Delta_L}{H^2 + k_{\nu\nu^{\prime}}^2
    \Delta_L^2} \ ,
\end{equation}
where
\begin{equation}
\label{xinunu'}
    k_{\nu\nu^{\prime}} = \frac{2}{N+1} \sum_{n=1}^N
    \left| \sin \left( \frac{\pi \nu n}{N+1} \right)
    \sin \left( \frac{\pi \nu^{\prime} n}{N+1} \right) \right | \ .
\end{equation}
\label{P(H)Lorentzian}
\end{subequations}
As is seen, the distribution function of $H_{\nu\nu^{\prime}}$ is
also a Lorentzian. For the diagonal elements, $k_{\nu\nu} = 1$,
independently of $\nu$, i.e., $P_{\nu\nu}(H)$ has the same width as
the bare distribution~(\ref{Lorentz}), which implies that there is
no exchange narrowing in this case. The off-diagonal elements are
distributed differently, depending on $\nu$ and $\nu^{\prime}$. For
our purpose, namely theoretically estimating the localization length
in the neighborhood of the lower band edge, only $k_{12}$ is
relevant (see below). The analytical result reads $k_{12} =
8/(3\pi)$, which does not show an exchange narrowing effect either.
For arbitrary $\nu$ and $\nu^{\prime}$ the sum in
Eq.~(\ref{xinunu'}) can not be evaluated analytically. However, it
can be seen that for large $N$ it scales linearly with $N$. Thus, in
the limit of large $N$ the exchange narrowing effect is absent for
all $H_{\nu\nu^{\prime}}$.

\subsection{Estimates of the localization length}
\label{Theoretical estimates}

The expressions in the preceding section are valid generally.
However, only in the perturbative limit, when $H_{\nu\nu^{\prime}} <
|E_{\nu} - E_{\nu^{\prime}}|$, does it make sense to consider
$|\nu\rangle$ as the (approximate) eigenstates and to interpret the
widths of the distributions $P_{\nu\nu}(H)$ as linewidths for the
absorption peak of that state. If this inequality does not hold, the
off-diagonal matrix elements $H_{\nu\nu^{\prime}}$ mix the exciton
states, resulting in their localization on segments of the chain.
Obviously, the perturbative limit is never reached for infinite
chains, as then the energy separations of states adjacent in energy
get infinitesimally small. It has been shown, however, that the
above mixing arguments lead to an excellent estimate of the typical
exciton localization length near the bare band bottom if applied
self-consistently to states within finite localization
segments.\cite{Malyshev91,Malyshev95}

The self-consistent argumentation is valid for states localized on
chain segments much smaller than the chain length and much bigger
than a single molecule ($1 \ll N^* \ll N$). Two notions underly the
reasoning: (i) States localized on the same chain segment undergo
level repulsion, i.e., they have a finite energy difference. In
particular, the energy separation between the two bottom states in a
localization segment of length $N^*$ is approximately given by the
energy difference $E_2^*-E_1^*$, where $E_\nu^*$ is given by
Eq.~(\ref{Enu}) for $N=N^*$. Here we used the fact that the states
resemble those of a finite homogeneous chain with length $N^*$, in
particular in the sense that the lowest exciton state on the segment
has a wave function without nodes, while the next higher state has
one node.\cite{Malyshev95} (ii) Two homogeneous basis states $\nu$
and $\nu^{\prime}$ localized on a chain segment of length $N^*$ are
mixed by $H_{\nu,\nu^{\prime}}^*$ given by Eq.~(\ref{Hnunu'}) with
$N$ replaced by $N^*$. The central argument in estimating the
localization size near the band bottom is now that this size adjusts
itself such that $H^*_{21}=E_2^{*}-E_1^{*}$. Namely, if $H^*_{21} <
E_2^{*}-E_1^{*}$, the disorder would only be perturbative and the
states would increase their spread. On the other hand, if $H^*_{21}
> E_2^{*}-E_1^{*}$, the disorder would strongly mix the two states
and would localize the exciton wave functions further.

As we have seen above, for Gaussian disorder, or any other disorder
distribution with a bounded second moment, exchange narrowing of
$H_{21}$ takes place, such that its typical value is $H^*_{21} =
\Delta_G/\sqrt{N^*}$. Furthermore, assuming that $N^{*} \gg 1$ we
have $E_2^{*} - E_1^{*}=3\pi^2J/N^{*2}$. Thus, the requirement
$H^*_{21}=E_2^{*}-E_1^{*}$ leads to the
estimate~\cite{Malyshev91,Malyshev95}
\begin{equation}
\label{scaling1}
    N^{*} = \left(3\pi^2\frac{J}{\Delta_G}\right)^{2/3} \ .
\end{equation}
for the typical localization size. This power-law behavior is in
excellent agreement with previous numerical
calculations,~\cite{Schreiber82, Fidder91} as well as with the
analytical scaling relation obtained within the Coherent Potential
Approximation (CPA).~\cite{Boukahil90}

For systems with Lorentzian disorder, no exchange narrowing occurs,
i.e., $H^*_{21} = 8\Delta_L/(3\pi)$. Now the requirement
$H^*_{21}=E_2^{*}-E_1^{*}$ yields
\begin{equation}
\label{scaling2}
        N^{*} =
        \left(\frac{9\pi^3}{8}\frac{J}{\Delta_L}\right)^{1/2},
\end{equation}
which reveals a different power-law scaling than for Gaussian
disorder. In Sec.~\ref{sec:scaling} we will find that the
$\Delta_L^{-1/2}$ scaling indeed agrees with numerical results.

While our main interest is in the optically dominant band-edge
states, it is interesting to apply the above arguments also to the
band center, and compare to previous
results.~\cite{Kramer93,Fyodorov92}  Near the band center, $E_\nu^*
- E_{\nu^\prime}^* \propto 2\pi J/N^*$, with $N^*$ now indicating
the typical localization size at the band center. Equating this
quantity to the exchange narrowed disorder strength,
$\Delta_G/\sqrt{N^*}$, we find $N^* \propto (J/\Delta_G)^2$. This
indeed is the well known disorder scaling of the localization length
in one-dimensional systems with Gaussian site
disorder,~\cite{Kramer93} which in Ref. \onlinecite{Fyodorov92} has
also been obtained by an analytical calculation of the inverse
participation ratio, performed within the framework of a
one-dimensional nonlinear supermatrix $\sigma$ model.

\section{Hidden spectral structure for Lorentzian disorder}
\label{sec:wavefunctions}

As we mentioned above, the optically dominant states in a disordered
exciton chain with negative transfer interactions occur in the
neighborhood of the band bottom, predominantly in the Lifshits tail.
The exciton states in this tail exhibit a hidden
structure,~\cite{Malyshev95} where doublets of $s$- and $p$-like
states often occur on the same localization segments (and more
rarely, triplets). As a result the low-temperature optical response
of the chain behaves approximately as that of a collection of
segments with typical size $N^*$ equal to the typical localization
length of the optically dominant states. More detailed statistics
were obtained in Ref.~\onlinecite{Malyshev01}, where the disorder
scaling of the localization length, the oscillator strength, and the
energy spacings $E_2^* - E_1^*$ were scrutinized numerically for
Gaussian diagonal disorder. In this section, we report the results
of similar numerical calculations with Lorentzian diagonal disorder
and show that this both alters the hidden structure near the bare
band bottom and induces a relatively high density of strongly
localized low energy wave functions.

\begin{figure}[ht]
\centerline{\includegraphics[width=0.8\columnwidth,clip]{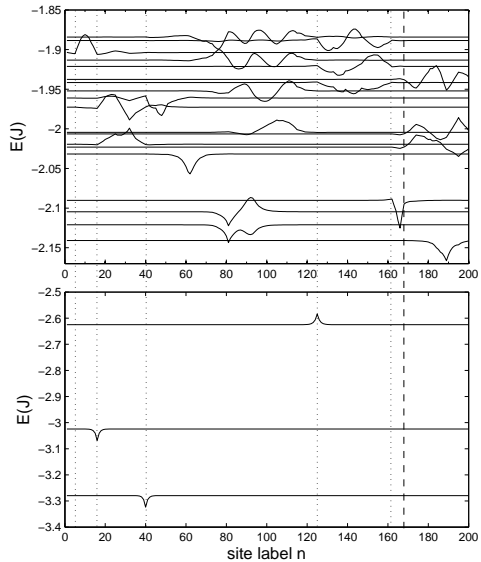}}
    \caption{A realization of the low energy wave functions near
    the band bottom (upper panel) and well below it (lower panel)
    for a chain of $N=200$ sites and a Lorentzian distribution of
    site energies (HWHM $\Delta_L=0.05J$). In the upper panel
    three types of states are observed: (i) Multiplets of states
    (upper panel), (ii) states strongly localized
    on short segments, and (iii) states that form a
    conventional hidden level structure (see text for details).
    The solid vertical
    lines indicate the sites with large negative energy fluctuations, around
    which states deep in the tail are localized (lower panel): $n=5$, $n=16$,
    $n=40$, $n=125$, and $n=162$, while the dashed vertical line
    corresponds to the high energy site $n=168$. Two low energy
    exciton states (at $n=5$ and $n=162$) are not shown.
    }
\label{fig:multi}
\end{figure}

The exciton wave functions for a given realization of the disorder
are straightforwardly obtained by diagonalizing the Hamiltonian
Eq.~(\ref{ham}), where we pick the site energies $E_n$ from a
Lorentzian distribution Eq.~(\ref{Lorentz}). In
Fig.~\ref{fig:multi}, we show a typical realization of the wave
functions in the neighborhood of the bare exciton band edge $E_b = -
2J$ (upper panel) and well below it (lower panel), calculated for
$\Delta_L = 0.05 J$. It is clearly seen that the localization
properties of these two subsets of wave functions differ
substantially. The states near the band edge are much more extended
than those deep in the tail and exhibit a hidden multiplet
structure, similar to the case of Gaussian
disorder.~\cite{Malyshev95,Malyshev01} The important difference is
that for Gaussian disorder, one usually finds only singlets and
doublets of states localized on a particular chain segment. In the
case of a Lorentzian distribution, on the other hand, we see the
occurrence of multiplets of three and even four states, often on
segments with sharply defined boundaries. It should be stressed that
this situation is typical; looking at various realizations, we
always found higher order multiplets. Below, we provide an
explanation for this special property of Lorentzian disorder.

As we mentioned above, the states deep in the tail of the DOS (lower
panel in Fig.~\ref{fig:multi}) look different from those in the
neighborhood of the band edge. They are, first, localized much more
strongly (in fact, on a few sites, as can be seen by the eye and is
confirmed by the participation number) and they are always
represented by $s$-like singlets. These states originate from a
large negative fluctuation of one particular site energy. For
moderate disorder magnitudes, such fluctuations occur frequently for
Lorentzian disorder, in contrast to Gaussian disorder.

Such outliers in energy have another consequence: they act as
natural barriers, providing a segmentation of the chain into smaller
subchains. In the realization of Fig.~\ref{fig:multi}, these
outliers occur at the positions $n=5$, $n=16$, $n=40$, $n=125$,
$n=162$ and $n=168$. Segments between such barriers that happen to
have a length of around the typical localization length $N^*$ of the
band-edge states (upper panel), can easily support the formation of
multiplets. For instance, a triplet of localized states occurs
between $n = 125$ and $n = 162$, and a quartet is observed between
$n = 168$ and $n = 200$. On the other hand, more strongly localized
states may occur near the band edge as well. This happens if two
closely spaced sites acquire large energy fluctuations, creating a
segment considerably smaller than $N^*$. The states localized
between $n = 5$ and $n = 16$ and between $n = 162$ and $n = 168$
represent two examples of this second type of exciton state.
Finally, the segments that are appreciably larger than $N^*$ show a
hidden level structure similar to that for Gaussian disorder, that
is, where the localization segments have poorly defined boundaries
and can be seen to overlap each other. This third type of exciton
state can be seen in the segment between $n=40$ and $n=125$ in
Fig.~\ref{fig:multi}.

We proceed with a brief analysis of the wave functions of the
exciton states deep in the tail of the DOS, which coincide with
(some of) the aforementioned segment boundaries. These strongly
localized states are centered around very low energy sites, and
typically have localization lengths, as defined by the participation
ratio in Eq.~(\ref{part}), of around a few sites. The wave functions
decay approximately exponentially when one moves away from the
central low energy site.

A simple model for these wave functions is given by a particle of
mass $m$ moving in a $\delta$-function potential well. It is well
known that such a well supports one bound state (with negative
energy $E$), the wave function of which behaves proportionally to
$\exp(-|x|/\xi)$ (see, e.g., Ref.~\onlinecite{Griffiths95}). The
relation between the penetration depth $\xi$ and the energy is given
by~\cite{Griffiths95} \begin{equation}\label{xi}
    \xi=\sqrt{\frac{\hbar^2}{m}}\left(-E\right)^{-1/2} \ .
\end{equation}
Clearly, the meaning of $\xi$ is similar to the localization length
$N^*$; both are measures of the extent of the wave function. The
localization length calculated through the participation number Eq.
\ref{part} is more suited for wave functions of unknown, arbitrary
shapes. The known shape of the wave functions for the low-energy
states, however, suggests to consider their exponential decay
lengths and investigate whether their energy scaling resembles Eq.
\ref{xi}.

\begin{figure}[ht]
\centerline{\includegraphics[width=\columnwidth]{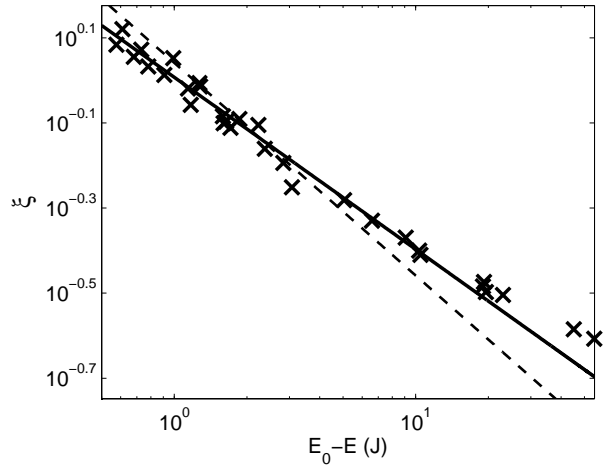}}
    \caption{Double logarithmic plot of the numerically obtained
    penetration depth $\xi$ (in lattice units) as a function of
    energy $E_0-E$. The expected decay of the penetration depth
    with energy is clearly seen. The solid line shows a power
    law fit, while the dashed line represents the $\xi \propto
    \left(E_0-E\right)^{-1/2}$ behavior suggested by the
    $\delta$-function potential model.
    }
\label{fig:xi}
\end{figure}

As was shown in Refs.~\onlinecite{Klugkist08}
and~\onlinecite{Klugkist07}, in the neighborhood of the bare exciton
band edge, a universal energy variable $\tilde{\varepsilon}$ exists,
which is given by
\begin{equation}
    \tilde{\varepsilon}=\frac{\varepsilon -
    \varepsilon_b + a\sigma^{\alpha}}{b\sigma^{\alpha}} \ ,
\end{equation}
where $\varepsilon = E/J$, $\varepsilon_b = E_b/J$ is the bare
exciton band edge energy ($\varepsilon_b=-2$ for nearest-neighbor
interactions), and the disorder strength $\sigma$ is either
$\Delta_G/J$ or $\Delta_L/J$ for Gaussian and Lorentzian disorder,
respectively. In terms of $\tilde{\varepsilon}$, functions of
energy, like the DOS, become universal. The numerical coefficients
$a$, $b$, and $\alpha$ depend on the type of disorder considered as
well as on whether we include only nearest-neighbor or all
dipole-dipole interactions. In particular, for nearest-neighbor
interactions and Gaussian diagonal disorder, this set is given by
$a=0$, $b=1$, and $\alpha=4/3$,~\cite{Klugkist07} while in the case
of nearest-neighbor interactions and Lorentzian diagonal disorder,
we should use $a=1$, $b=4$ and $\alpha=1$.~\cite{Klugkist07} We
define a reference energy $E_0$ that is dependent on the disorder
magnitude, which is the energy that corresponds to
$\tilde{\varepsilon}=0$ on the universal energy scale.

An analysis of the numerically determined wave functions indeed
confirms the expected general trend that the lower energy states are
localized more strongly than the band edge states. For the deep tail
energy states, the penetration depth $\xi$ was calculated by fitting
an exponential to the numerically determined wave function. The
result is plotted in Fig. \ref{fig:xi} against the state's energy
measured from the reference energy $E_0$. We observe that these data
approximately obey a power law, although not with quite the same
exponent as suggested by Eq.~(\ref{xi}). The best fit was obtained
using the power law $\xi(E) = 1.017
\left[(E_0-E)/J\right]^{-0.405}$. The deviation of the numerically
determined exponent from the estimated power law is most likely
caused by the fact that the $\delta$-function model assumes a
continuous position variable and an infinitely narrow and infinitely
deep well, while our model Hamiltonian is discrete and can thus only
approximately behave as in the $\delta$-function model. We also note
that the very low energy states are very symmetric, while the
exciton states that are closer to the band edge show more asymmetry;
this makes sense, since for the more energetic states the
surrounding sites become more important, at the expense of the
central low-energy site. As the energies of the surrounding sites
are random, this leads to the observed asymmetry.

Summarizing this section, we point out that the nature of the
localization in the Lifshits tail of the DOS for Lorentzian disorder
clearly differs from the usual Gaussian model in a number of
aspects, in particular, it alters the hidden level structure and
introduces strongly localized states deep in the tail which are far
less likely to occur for Gaussian stochastic variables. Close to the
bare exciton band edge, we observe three types of states: strongly
localized $s$-like states on segments much smaller than the
localization length $N^*$, multiplets on segments of lengths
comparable to $N^*$, and finally, states that resemble the
conventional hidden level structure that is also found for Gaussian
disorder. In the next section we provide a more detailed analysis of
the scaling of the localization length for both types of disorder in
the neighborhood of the bare exciton band edge.

\section{Localization length distributions and scaling}
\label{sec:scaling}

The localization length of a given exciton state $|\nu \rangle$ can
be calculated numerically using the participation number $N_{\nu}$,
Eq.~(\ref{part}). It is a fluctuating quantity and is thus more
accurately described by analyzing its probability distribution. We
focus on the energy range where the optically dominant states
reside, i.e., in the neighborhood of the bare exciton band edge,
$E_b = -2J$, and predominantly, just below it. To ensure a fair
comparison, we fix the endpoints of the energy interval under
consideration in terms of the universal energy variable mentioned in
Section \ref{sec:wavefunctions}. In the following simulations, it
ranges from $\tilde{\varepsilon}_i=-0.1$ to
$\tilde{\varepsilon}_f=0$.

\begin{figure}[ht]
\centerline{\includegraphics[width=\columnwidth]{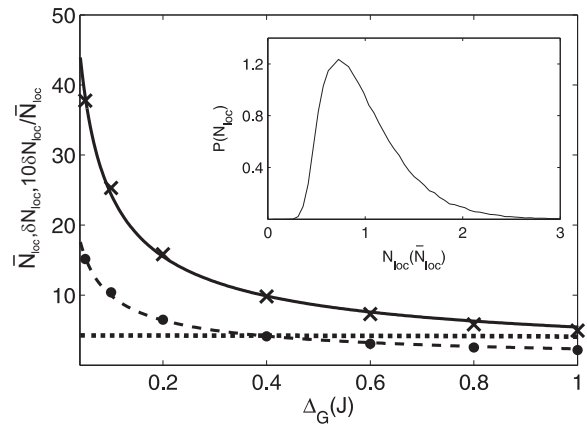}}
    \caption{Disorder scaling of the average localization length
    (denoted by crosses) and its standard deviation of the distribution
    (denoted by the dots) for a chain of length $N=250$ sites with
     Gaussian disorder. Power law fits for both have been included:
     the solid line corresponds to
     $\bar{N}_{loc}=5.43\left(\Delta_G/J\right)^{-0.65}$, and the
     dashed line gives $\delta N_{loc}=2.33\left(\Delta_G/J\right)^{-0.63}$.
     Additionally, the quotient of the standard deviation and the
     average localization length is shown by the dotted line
     (multiplied by a factor of 10 for better visibility), which
     clearly shows this ratio to be constant to a good approximation.
     In the inset, we show the localization length distribution for
     $\Delta_G=0.2J$, plotted on a $N_{loc}$ scale which is normalized
     to $\bar{N}_{loc}$. The resulting function is universal, as it
     does not depend on $\Delta_G$ (see text).
    }
\label{fig:scaling1}
\end{figure}

\begin{figure}[ht]
\centerline{\includegraphics[width=\columnwidth]{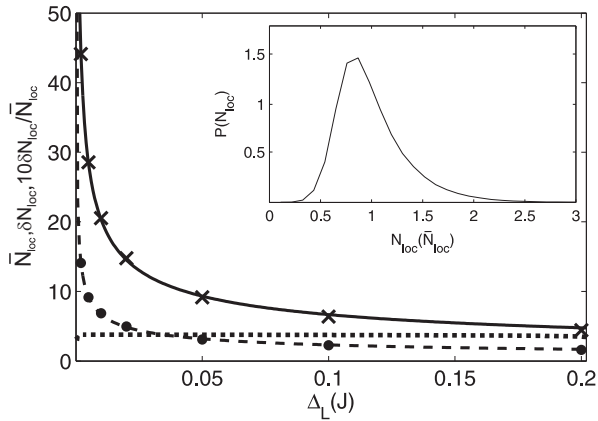}}
    \caption{Same as in Fig.~\ref{fig:scaling1}, but now for
    Lorentzian disorder with HWHM $\Delta_L$. The power law fits for
    the average localization length (solid line) and the standard
    deviation (dashed line) shown in the figure correspond to
    $\bar{N}_{loc}=2.19\left(\Delta_L/J\right)^{-0.48}$ and
    $\delta N_{loc}=0.80\left(\Delta_L/J\right)^{-0.46}$,
    respectively. The distribution in the inset was obtained
     for $\Delta_L=0.05J$.
    }
\label{fig:scaling2}
\end{figure}

The probability distribution of the localization length collected in
this energy region,
$P(N_{loc})=\sum_{\nu}^{\prime}\delta(N_{loc}-N_{\nu})$ (the prime
restricting the summation to the selected energy interval), are
plotted in the insets of Fig.~\ref{fig:scaling1} (Gaussian disorder
with $\Delta_G=0.2J$) and Fig.~\ref{fig:scaling2} (Lorentzian
disorder with $\Delta_L=0.05J$). They show an asymmetric shape,
similar to the one presented in Ref.~\onlinecite{Malyshev01}. The
asymmetry is caused  by the fact that on average, the states with
smaller localization lengths reside deeper in the DOS tail, where
the DOS itself is small. If we instead go towards the exciton band
edge, the situation is reversed.

Analyzing the moments of the various distributions indeed highlights
the difference between the localization behavior for the two types
of disorder, as predicted in Sec. \ref{sec:background}. In
particular, for Gaussian disorder, we find
$\bar{N}_{loc}=5.43\left(\Delta_G/J\right)^{-0.65}$, which has an
exponent that is in good agreement with the theoretical power law
Eq. (\ref{scaling1}). Lorentzian disorder, on the other hand, leads
to the scaling with a different exponent, 
$\bar{N}_{loc}=2.19\left(\Delta_L/J\right)^{-0.48}$, again in good
agreement with the theoretical estimate of Eq.~(\ref{scaling2}),
$N^{*} \approx 5.91 \left(\Delta_L/J\right)^{-1/2}$. There is a
deviation in the numerical prefactors, which is not surprising given
the arbitrariness in defining the localization length. Most
important are the different powers found for both types of disorder,
owing to the absence of exchange narrowing for Lorentzian disorder.

It is remarkable that for both types of disorder, the standard
deviation is observed to scale with almost the same exponent as the
first moment. While the disorder amplitude varies over two orders of
magnitude, the standard deviation divided by the first moment
changes by less than ten percent (see Figs.~\ref{fig:scaling1}
and~\ref{fig:scaling2}). This implies that on the scale of the first
moment the localization length distributions have universal shapes,
which are presented in the insets in Figs.~\ref{fig:scaling1}
and~\ref{fig:scaling2}. The Lorentzian disorder model is observed to
yield a narrower distribution.

All the results presented above were obtained using nearest-neighbor
interactions. Obviously, a similar analysis can be performed when
accounting for all dipole-dipole interactions. Doing so induces a
shift in the band bottom ($\varepsilon_b=-2.404$ in this case) and
modifies the DOS in the Lifshits tail.~\cite{Fidder91} In turn, this
changes the exponents in disorder scalings,~\cite{Fidder91,
Malyshev01} however it does not influence the physics that is
responsible for the differences in localization properties of
Gaussian and Lorentzian disorder.

\section{Summary}
\label{sec:summary}

In this study, we performed a comparison of the localization
properties of a one-dimensional Frenkel exciton model with Gaussian
and Lorentzian uncorrelated diagonal disorder, with a special focus
on the energy region below the bare exciton band edge. We have found
that the divergent second and higher moments of Lorentzian
distributions lead to a number of interesting modifications of the
localization behavior as compared to Gaussian disorder. A striking
example is the absence of exchange narrowing, which, as we have
shown, results in a different disorder scaling of the localization
length from the one obtained for Gaussian disorder. This theoretical
prediction is supported by numerical calculations, which reveal
power law scalings with exponents that are in excellent agreement
with the theoretical estimates. Moreover, the standard deviation of
the localization length distribution scales with disorder with
almost the same exponent as the average localization length,
implying that the shape of the distribution is universal.

We have also shown that the wave functions in a chain with
Lorentzian disorder have a hidden structure that differs
substantially from the one found for Gaussian disorder. First of
all, Lorentzian disorder gives rise to a relatively high density of
strongly localized exciton states deep in the DOS tail. They
resemble the bound states of a $\delta$-function potential and occur
as a result of large fluctuations of certain monomer transition
energies; this is a special properety of Lorentzian distributions,
resulting from the fact that their second and higher moments
diverge. These exciton wave functions are localized on only a few
monomers; their spatial extent decreases with decreasing energy. The
more extended states in the Lifshits tail, close to the band bottom,
form manifolds of states localized on various segments of the chain.
Lorentzian disorder produces a higher amount of multiplets as
compared to Gaussian disorder, where only singlets and doublets are
commonly encountered. In addition, we have also shown that strongly
localized $s$-like singlet states may occur near the band edge on
segments that are appreciably smaller than the localization length
$N^*$, while an exciton level structure that is comparable to the
one for a Gaussian disorder model is found for segments that are
considerably larger than the localization length. These changes may
result in different behavior of the temperature-dependent energy
transport in J-aggregates for these two types of disorder.

\acknowledgments

We thank A. Eisfeld and J. S. Briggs for inspiring discussions on
Lorentzian disorder.

\appendix

\section{Dipolar interactions as a source of Lorentzian
disorder}
\label{non Gaussian disorder}

In this Appendix, we show that random dipolar interactions may
result in Lorentzian disorder. Consider a (transition) dipole
$d_0{\bf l}_0$ surrounded by other dipoles $d_1{\bf l}_i$ (all of
the same magnitude), where $d_0$ and $d_1$ are their magnitudes,
while ${\bf l}_0$ and ${\bf l}_i$ denote their orientations. We
assume that the surrounding dipoles are randomly distributed in a
volume ${\cal V}$ and also oriented within the solid angle $4\pi$
according to a probability density $f({\bf l}_i)$, so that the joint
probability density to find any of them somewhere in space and
somehow oriented is $f({\bf l}_i)/{\cal V}$. The object of our
interest is the probability distribution of the total dipole-dipole
interaction of the central dipole $d_0{\bf l}_0$ with the
surrounding dipoles $d_1{\bf l}_i$.
\begin{subequations}
\begin{equation}
    V = \sum_{i=1}^{\cal N} \frac{V_{0}}{r_i^3}
    \phi({\bf l}_0,{\bf l}_i,{\bf n}_i) \ ,
 \label{Interaction}
\end{equation}
\begin{equation}
    \phi({\bf l}_0,{\bf l}_i,{\bf n}_i) = {\bf l}_0{\bf l}_i -
    3 ({\bf l}_0{\bf n}_i) ({\bf l}_i {\bf n}_i) \ ,
\end{equation}
\end{subequations}
where $V_0 = d_0d_1/a^3$ is a constant, $r_i$ is the (dimensionless)
distance between the dipoles $d_0{\bf l}_0$ and $d_1{\bf l}_i$, and
$\phi({\bf l},{\bf l}_i,{\bf n}_i)$ is the orientational factor with
${\bf n}_i$ the unit vector along ${\bf r}_i$. The quantity $V$ can
be associated with the so-called solvent-induced shift of the
monomer transition energy in molecular aggregates (see, e.g., the
textbooks~\onlinecite{Davydov71} and~\onlinecite{Agranovich82}).

The probability distribution of $V$ reads
\begin{equation}
 \label{V distribution}
    P(V) = \left\langle \delta \left(V - \sum_{1=1}^{\cal N}
    \frac{V_0}{r_i^3} \phi({\bf l}_0,{\bf l}_i,{\bf n}_i)
    \right) \right\rangle \ ,
\end{equation}
where the angular brackets denote the average over the distribution
of surrounding dipoles,
\begin{equation}
\left\langle ...\right\rangle =
    \prod_{i=1}^{\cal N} \int \frac{d{\bf r}_i}{\cal V}
    \int d{\bf l}_if({\bf l}_i)....
\end{equation}
Furthermore, using the integral representation for the
$\delta$-function and performing the average in Eq.~(\ref{V
distribution}), we obtain
\begin{widetext}
\begin{equation}
 \label{After averaging}
    P(V) = \frac{1}{2\pi} \int_{-\infty}^{\infty} dt\, e^{iVt}\left [
    \int \frac{d{\bf r}}{{\cal V}} \int d{\bf l}_if({\bf l}_i)
    \exp\left( -i\frac{V_0t}{r^{3}} \phi \right) \right]^{\cal N}
\nonumber\\
\end{equation}
\begin{equation}
     = \frac{1}{2\pi} \int_{-\infty}^{\infty} dt\, e^{iVt} \left\{
    1 - \frac{n_0}{{\cal N}}\int d{\bf r} \int d{\bf l}_if({\bf
    l}_i)
    \left[ 1 - \exp\left( -i\frac{V_0t}{r^{3}} \phi
    \right) \right] \right\}^{\cal N}
\end{equation}
\end{widetext}
where $n_0 = {\cal N}/{\cal V}$ is the number density of the
surrounding dipoles. In the thermodynamic limit ${\cal N} \to
\infty$, ${\cal V} \to \infty$, while $n_0 = {\cal N}/{\cal V} =
\mathrm{const}$, Eq.~(\ref{After averaging}) reduces to
\begin{equation}
 \label{P(V)}
    P(V) = \frac{1}{2\pi} \int_0^{\infty}dt \, e^{iVt}
    \exp\left[ -\frac{4\pi}{3}\xi n_0 V_0 t \right] +
    \mathrm{c.c.}\ .
\end{equation}
Here, $\xi = \int_0^{\infty} dz\, z^{-2} \int d{\bf n} \int d{\bf
l}f({\bf l}_i)[1 - \exp(-iz\phi)]$ and we made the substitution $z =
V_0t/r^3$. The integral in Eq.~(\ref{P(V)}) can be evaluated
analytically. The result is a Lorentzian distribution
\begin{equation}
\label{Lorentzian}
    P(V) = \frac{1}{\pi} \frac{\Gamma}{(V - V_0)^2 + \Gamma^2}
\end{equation}
shifted from zero by $V_0 = (4\pi/3) $\text{Im}$\xi \, n_0V_0$ and
having a HWHM $\Gamma = (4\pi/3) $\text{Re}$\xi \, n_0V_0$. Both
magnitudes are determined by the dipole-dipole interaction at the
average distance between dipoles, $(4\pi/3)n_0V_0$

\section{Derivations of Eqs.~\ref{P(H) Gaissian} and \ref{P(H)Lorentzian}}
\label{Exchange narrowing}

Here, we present the derivation of the probability distribution for
the matrix elements $H_{\nu \nu^{\prime}}$ given by
Eq.~(\ref{Hnunu'}). By definition,
\begin{equation}
\label{P(H) definition}
    P_{\nu\nu^{\prime}}(H)
     =  \left \langle \delta ({H - H_{\nu\nu^{\prime}}}) \right\rangle
     =  \frac{1}{2\pi} \int_{-\infty}^{\infty} \, dt e^{iHt}
    \left \langle e^{-iH_{\nu\nu^{\prime}}t} \right \rangle \ ,
\end{equation}
where the angular brackets $\left<...\right> = \prod_{n=1}^N \int
dE_n p(E_n) ....$ denote the average over disorder realizations,
with $p(E_n)$ being either a Lorentzian or a Gaussian distribution
function. Further, we use for $p(E_n)$ a representation through the
characteristic function:
\begin{equation}
\label{Characteristic function representation}
    p(E_n) = \frac{1}{2\pi} \int_{-\infty}^{\infty} dt_n
    e^{iE_nt_n}\exp(-|\Delta t_n|^{\alpha}) \ .
\end{equation}
For $\alpha = 1$, this formula gives a Lorentzian with HWHM equal to
$\Delta$, while for $\alpha = 2$ we get a Gaussian with standard
deviation $\Delta$. Using this representation in Eq.~(\ref{P(H)
definition}), we obtain
\begin{widetext}
\begin{equation}
    P_{\nu \nu^{\prime}}(H) =  \frac{1}{2\pi}\int_{-\infty}^{\infty}dt
    \> e^{iH t}
    \prod_{n=1}^{N}  \int_{-\infty}^{\infty}dt_n
    \exp(-|\Delta_{\alpha} t_n|^{\alpha}) \>
    \frac{1}{2\pi}\int_{-\infty}^{\infty}dE_n \> e^{iE_n[t_n-t\beta_{\nu
    \nu^{\prime}}(n)]}
\end{equation}
\end{widetext}
with
\begin{equation}
\label{beta}
    \beta_{\nu \nu^{\prime}}(n)
    = \frac{2}{N+1} \sin \frac{\pi \nu n}{N+1}
    \sin \frac{\pi \nu^{\prime} n}{N+1} \ .
\end{equation}
The integral over $E_n$ yields a $\delta$-function, which finally
gives us
\begin{equation}
\label{P}
    P_{\nu \nu^{\prime}}(H) =  \frac{1}{2\pi}\int_{-\infty}^{\infty}dt \> e^{iH t}
    \exp\left(-\left|k_{\nu \nu^{\prime}}\Delta t\right|^{\alpha}
    \right)
    \ ,
\end{equation}
where
\begin{equation}
\label{knunu'}
    k_{\nu \nu^{\prime}} =
    \left(\sum_{n=1}^N\left|\beta_{\nu \nu^{\prime}}(n)\right|^{\alpha}\right)^{1/\alpha} \
    .
\end{equation}
As is seen from Eq.~(\ref{P}), the distribution $P_{\nu
\nu^{\prime}}(H)$ is a Lorentzian and a Gaussian for $\alpha = 1$
and $\alpha = 2$, respectively, only with renormalized HWHM and
standard deviation, given by $k_{\nu \nu^{\prime}}\Delta$.

For $\alpha =2$, the summation in Eq.~(\ref{knunu'}) can be
performed analytically for any $\nu$ and $\nu^{\prime}$. The result
reads: $k_{\nu \nu} = 1/\sqrt{(3/2)(N + 1)}$ and $k_{\nu
\nu^{\prime}} = 1/\sqrt{N + 1}$ for $\nu \ne \nu^{\prime}$. In the
case of $\alpha = 1$, we get $k_{\nu \nu} = 1$, while at $\nu \ne
\nu^{\prime}$ the sum in Eq.~(\ref{knunu'}) depends on $\nu$ and
$\nu^{\prime}$ and does not have a simple expression.

\end{document}